\documentstyle[11pt]{article}

\newcounter{subequation}[equation]

\makeatletter
\newenvironment{subeqnarray}
  {\arraycolsep1pt
    \def\@eqnnum\stepcounter##1{\stepcounter{subequation}{\reset@font\rm
      (\theequation\alph{subequation})}}\eqnarray}
  {\endeqnarray\stepcounter{equation}}
\makeatother

\newcounter{statement}
\newenvironment{statement}[4]
  {\par\refstepcounter{statement}
    \noindent#1#2 \arabic{statement} #4\unskip: #3}{\par\vspace{2mm}}
\newenvironment{Lemm}
  {\begin{statement}{\bf}{Lemma}{\sl}}{\end{statement}}
\newenvironment{Prop}
  {\begin{statement}{\bf}{Proposition}{\sl}}{\end{statement}}
\newenvironment{statement*}[4]
  {\par\noindent#1#2 #4\unskip: #3}{\par\vspace{2mm}}
\newenvironment{Coro}
  {\begin{statement*}{\bf}{Corollary}{\sl}}{\end{statement*}}
\newenvironment{Prf}
  {\begin{statement*}{\sl}{Proof}{\rm}}{\end{statement*}}

\begin{document}

\hbox to\hsize{%
  \vbox{\hbox{Submitted to}\hbox{\sl Commun.\ Math.\ Phys.}}\hfil
  \vbox{\hbox{MPI-PhT/94-20}\hbox{May 1994}}}
\vspace{1cm}
\begin{center}
{\LARGE On the Limiting Solution of the\\
        Bartnik-McKinnon Family}\\[5mm]
{\large Peter Breitenlohner {\normalsize and} Dieter Maison}\\ [3mm]
{\small\sl Max-Planck-Institut f\"{u}r Physik} \\
{\small\sl --- Werner Heisenberg Institut ---} \\
{\small\sl F\"{o}hringer Ring 6}\\
{\small\sl 80805 Munich (Fed. Rep. Germany)}\\[1cm]
{\bf Abstract}
\end{center}
\begingroup \addtolength{\leftskip}{1cm} \addtolength{\rightskip}{1cm}
\noindent
We analyze the limiting solution of the Bartnik-McKinnon family
and show that its exterior is an extremal Reissner-Nordstr{\o}m black
hole and not a new type of non-abelian black hole as claimed in a
recent article by Smoller and Wasserman.

\endgroup
\vspace{1cm}

\noindent
The purpose of this short communication is to correct some erroneous
statements made in a recent article by J.A.~Smoller and A.G.~Wasserman
\cite{SW}.  This article concerns the limiting behaviour of an
infinite discrete family of regular, static, spherically symmetric
solutions of the Einstein-Yang-Mills equations (gauge group {\sl
SU(2)\/}), whose first few members were discovered by Bartnik and
McKinnon \cite{Bart}.  A general existence proof for this family was
given by Smoller and Wasserman \cite{Smoller} and by the present
authors together with P.~Forg\'acs \cite{BFM}.

In their article \cite{SW} the authors claim that a suitable
subsequence of the infinite family converges to some limiting solution
for all values of the radial coordinate $r\neq 1$.  The part of this
limit defined for $r>1$ is interpreted as a new type of black hole
solution with event horizon at $r=1$.  According to their claim the
function $W(r)$ parametrizing the Yang-Mills potential is non-trivial,
i.e., $W\not\equiv 0$ and tends to $+1$ or $-1$ for $r\to\infty$.  In
contrast we claim that the limiting solution for $r>1$ is given by the
extremal Reissner-Nordstr{\o}m (RN) solution with $W\equiv 0$.  This can
be easily derived from the results of our article \cite{BFM} and is
also strongly supported by numerical calculations.  Subsequently we
shall give a proof of this claim using the results of \cite{BFM}.

First we recall some definitions and results of \cite{BFM}.  The
variables $T$, $A$, $\mu$, $w$, and $\lambda$ used in
\cite{SW,Smoller} correspond to the quantities $(AN)^{-1}$, $\mu$,
$2m$, $W$, and $2b$ in \cite{BFM} and in this article.  We parametrize
the line element in the form
\begin{equation}\label{metra}
ds^2=A^2(r)\mu(r)dt^2-{dr^2\over\mu(r)}
  -r^2d\Omega^2\;,
\end{equation}
and use the `Abelian gauge'
\begin{equation}\label{Ans}
W_\mu^a T_a dx^\mu=
  W(r) (T_1 d\theta+T_2\sin\theta d\varphi) + T_3 \cos\theta d\varphi\;,
\end{equation}
for the static, spherically symmetric {\sl SU(2)\/} Yang-Mills field.

The field equations for $A$, $\mu$, and $W$ (see, e.g., Eqs.~(6)
in~\cite{BFM}) are singular at $r=0$ and $r=\infty$ as well as for
$\mu(r)=0$.  In order to desingularize them when $\mu\to0$ we introduce
$N=\sqrt{\mu}$, $U=NW'$, a new independent variable $\tau$ (with
$\dot{}=d/d\tau$), and $\kappa=(\ln rAN)\dot{}$ as additional dependent
variable.  The field eqs.\ are then equivalent to the autonomous first
order system
\begin{subeqnarray}\label{taueq}
  \dot r&=&rN\;,\\
  \dot W&=&rU\;,\\
  \dot U&=&{W(W^2-1)\over r}-(\kappa-N)U\;,\\
  \dot N&=&(\kappa-N)N-2U^2\;,\\
  \dot\kappa&=&1+2U^2-\kappa^2\;,\\
  (AN)\dot{}&=&(\kappa-N)AN\;,
\end{subeqnarray}
subject to the constraint
\begin{equation}\label{kappeq}
2\kappa N=1+N^2+2U^2-(W^2-1)^2/r^2\;.
\end{equation}
If the initial data satisfy this constraint then it remains true for all
$\tau$.

There exists a one-parameter family of local solutions with regular
origin where $W(r)=1-br^2+O(r^4)$, $\mu(r)=1+O(r^2)$ such that $W(r)$
and $\mu(r)$ are analytic in $r$ and $b$.  If we adjust $\tau$ such that
$\tau=\ln r+O(r^2)$ we obtain a one-parameter family of local solutions
of the system~(\ref{taueq}) which satisfy the constraint~(\ref{kappeq})
and are analytic in $\tau$ and $b$.

Similarly there exists a two-parameter family of local black hole
solutions with $W(r)=W_h+O(r-r_h)$, $\mu(r)=O(r-r_h)$ such that $W(r)$
and $\mu(r)$ are analytic in $r$, $r_h$, and $W_h$.  If we adjust $\tau$
such that $\tau=0$ at the horizon we obtain a two-parameter family of
solutions of (\ref{taueq},\ref{kappeq}) analytic in $\tau$, $r_h$, and
$W_h$ except for a simple pole in $\kappa(\tau)$ at the horizon.

Both types of initial data satisfy $\kappa\ge1$ and this relation
remains true for all $\tau$ due to the form of Eq.~(\ref{taueq}e).

In the following we exclude the case $W\equiv0$ and can therefore assume
$(W,U)\ne(0,0)$ for all (finite) $\tau$.  Integrating Eqs.~(\ref{taueq})
with regular initial data $r(\bar\tau)>0$, $N(\bar\tau)>0$,
$\kappa(\bar\tau)\ge1$ satisfying the constraint~(\ref{kappeq}) we
obtain solutions analytic for all $\tau>\bar\tau$ as long as
$N>-\infty$.  There are three possible cases:
\begin{enumerate}
\item[i)]
$N(\tau)$ has a zero at some $\tau=\tau_0$, the generic case.
Then
\begin{equation}\label{Wrbound}
(W^2(\tau_0)-1)^2=\left(1+2U^2(\tau_0)\right)r^2(\tau_0)\;,
\end{equation}
and $r$ has a maximum at $\tau=\tau_0$.
For $\tau>\tau_0$ we find that
$N<0$ and $r$, $W$, $U$, $\kappa$, $rN$, and $rAN$ remain analytic
at least as long as $r\ge0$.
\item[ii)]
$N(\tau)>0$ for all $\tau$ and $r(\tau)$ tends to infinity for
$\tau\to\infty$. These are the asymptotically flat solutions with
$(W,U,N,\kappa)\to(\pm1,0,1,1)$.
\item[iii)]
$N(\tau)>0$ for all $\tau$ and $r(\tau)$ remains bounded.
This is a new type of `oscillating' solution with
$(r,W,U,N,\kappa,A)\to(1,0,0,0,1,\infty)$ for $\tau\to\infty$ first
discussed in detail in \cite{BFM}.
\end{enumerate}

Analyzing the solutions with regular origin and their dependence on $b$
we have shown in \cite{BFM}:
\begin{enumerate}
\item[1.]
For each positive integer $n$ there exists a globally regular and
asymptotically flat solution with $n$ zeros of $W$ for at least one
value $b=b_n$ and there is at most a finite number of such values $b_n$.
\item[2.]
There exists an oscillating solution for at least one value $b=b_\infty$
and there is at most a finite number of such values $b_\infty$.
\item[3.]
The values $b_n$ have at least one accumulation point for $n\to\infty$
and each such accumulation point is one of the values $b_\infty$.
\end{enumerate}

\noindent
Completely analogous results hold for black hole solutions with fixed
$r_h<1$ and their dependence on $W_h$.

Let us analyze the oscillating solutions in some detail.
\gdef\bW{\overline{W}}
\gdef\bk{\overline{\kappa}}
Near the singular point $(r,W,U,N,\kappa)=(1,0,0,0,1)$ we introduce the
parametrization (with $\bW={W\over r}$ and $\bk=\kappa-1$)
\begin{subeqnarray}\label{Eqasy}
  \bW(\tau)&=&C_1 e^{-{1\over2}\tau}
  \sin({\sqrt{3}\over2}\tau+\theta)\;,\\
  U(\tau)&=&C_1 e^{-{1\over2}\tau}
  \sin({\sqrt{3}\over2}\tau+{2\pi\over3}+\theta)\;,\\
  N(\tau)&=&C_2 e^\tau+{2\over7}(\bW^2-U\bW+2U^2)\;,\\
  \bk(\tau)&=&C_4 e^{-2\tau}+\bW^2+2U\bW+2U^2\;,
\end{subeqnarray}
as in \cite{BFM} and compute $r$ from the constraint (\ref{kappeq})
\begin{equation}\label{req}
r^{-2}=\rho+\sqrt{\rho^2-\bW^4}\;,\quad
  {\rm where}\quad
  \rho={1\over2}(1-N)^2+\bW^2+U^2-\bk N\;.
\end{equation}
The functions $\theta$, $C_1$, $C_2$, and $C_4$ satisfy differential
eqs.
\begin{subeqnarray}\label{Eqdiff}
  \dot\theta&=&f_0\;,\\
  (C_1^2e^{-\tau})\dot{}&=&C_1^2e^{-\tau}(-1+f_1)\;,\\
  (C_2e^{\tau})\dot{}&=&C_2e^{\tau}+f_2\;,\\
  (C_4e^{-2\tau})\dot{}&=&-2C_4e^{-2\tau}+f_4\;,
\end{subeqnarray}
with `non-linear' terms $f_i$ that can be expressed as homogeneous
polynomials in $C_1^2e^{-\tau}$, $C_2e^{\tau}$, and $C_4e^{-2\tau}$ of
degree one for $f_0$ and $f_1$ and of degree two for $f_2$ and $f_4$
with $(r,\theta)$-dependent coefficients that are bounded as long as $r$
is bounded.

We can apply a general result for perturbed linear systems (see, e.g.,
\cite{Codd}~p.330) stating the existence of a stable manifold.  The
system~(\ref{Eqdiff}) has one unstable mode, $C_2e^{\tau}$, and hence
there exists a three-dimensional stable manifold of initial data, i.e.,
quadruples $Y=(\bW,U,N,\bk)$ such that $Y\to0$ for $\tau\to\infty$.
Eliminating the freedom to add a constant to $\tau$ we are left with a
two-parameter family of oscillating solutions.  In \cite{BFM} we have
derived the stronger result that $\theta$ and $C_1$ have a limit for
$\tau\to\infty$ (with $C_1(\infty)\ne0$) whereas $C_2e^{2\tau}\to0$ and
$C_4e^{-\tau}\to0$ for each member of this two-parameter family.
Consequently these oscillating solutions have infinitely many zeros of
$W$ and inifinitely many minima of $N$ as $r\to1$.

Conversely there exists a one-dimensional `unstable manifold' (i.e.,
stable manifold for decreasing $\tau$) of initial data such that $Y\to0$
for $\tau\to-\infty$.  These initial data $Y=(0,0,N,0)$ describe the
extremal RN black hole with $r=(1-N)^{-1}$.

In the following we analyze the behaviour of solutions for $b$ near (one
of the values) $b_\infty$ and in particular the behaviour of globally
regular solutions with $n$ zeros of $W$ in the limit $b_n\to b_\infty$
for $n\to\infty$.  In view of the analytic dependence of the solutions
on $b$ and $\tau$ the trajectories reach any given neighbourhood of the
singular point $Y=0$ for $b$ sufficiently close to $b_\infty$.
Trajectories missing the singular point cannot stay near it, they must
start to `run away'.  They will, however, remain close to the unstable
manifold.  In the limit $b_n\to b_\infty$ they converge to the unstable
manifold, i.e., extremal RN solution.

We can decompose $Y$ into its parts parallel and perpendicular to the
unstable manifold and measure the distance from the singular point $Y=0$
by
\gdef\NY{|Y|}
\gdef\Ypa{|Y_{\parallel}|}
\gdef\Ype{|Y_{\perp}|}
\begin{equation}\label{norm}
\NY=\max(\Ypa,\Ype)\;, \quad {\rm whith}
\quad \Ypa=|N|\;, \quad \Ype=\max(C_1^2e^{-\tau},|\bk|)\;.
\end{equation}

Using the distance function $|\cdot|$ we get from the smooth dependence
of the solutions on $b$ and $\tau$ that all solutions with $b\approx
b_\infty$ must come close to the singular point $Y=0$ for some
$\tau=\tau_0$.
\begin{Lemm}{}\label{Ldelta}
Given $b_\infty$ and any $\epsilon>0$ there exist some $\delta>0$ and
$\tau_0$ such that all solutions with $|b-b_\infty|<\delta$
satisfy $\NY(\tau_0)<\epsilon$ and $0<1-r(\tau_0)<\epsilon$.
\end{Lemm}

Let us analyze the behaviour of these trajectories in the neighbourhood
of $Y=0$.  The general result \cite{Codd} also states the existence of
some $\eta>0$ such that trajectories missing the singular point cannot
stay within $\NY<\eta$ for all $\tau$.  Due to the structure of
Eqs.~(\ref{taueq}), resp.~(\ref{Eqdiff}) this runaway is caused by the
growth of $N$.  The trajectories can therefore be characterized by three
possibilities:  They either run into the singular point $Y=0$ or miss it
on one or the other side; in the latter case either $N$ stays positive
and $r$ grows beyond $r=1$ or $N$ has a zero while $r<1$ and $r$ runs
back to $r=0$.  This is expressed by
\begin{Lemm}{}\label{Lthree}
There exists some $\eta>0$ such that for any solution of
Eqs.~(\ref{taueq}a--e,\ref{kappeq}) with $\NY<\epsilon\ll\eta$ and
$0<1-r<\epsilon$ at some $\tau=\tau_0$ there are three possible cases:
\begin{enumerate}
\item[a)]
$r<1$, $N>0$ for all $\tau>\tau_0$ and $Y\to0$ for $\tau\to\infty$,
\item[b)]
$r=1$ for some $\bar\tau>\tau_0$, $N=\eta$ for some $\tau_1>\bar\tau$,
$\dot N(\tau_1)>0$, and $0<N<\eta$, $\Ype<\epsilon$ for
$\tau_0<\tau<\tau_1$,
\item[c)]
$N=0$ for some $\bar\tau>\tau_0$, $N=-\eta$ for some $\tau_1>\bar\tau$,
$\dot N(\tau_1)<0$, and $r<1$, $|N|<\eta$, $\Ype<\epsilon$ for
$\tau_0<\tau<\tau_1$.
\end{enumerate}
\end{Lemm}
\begin{Prf}{}
The general result \cite{Codd} mentioned above states the existence of
some $\eta>0$ such that that either $Y\to0$ (case~{\bf a}) or $\NY=\eta$
for some $\tau_1$ (case~{\bf b} and~{\bf c}).  Choosing $\eta$ small
enough, Eq.~(\ref{Eqdiff}b) shows that the `amplitude'
$|C_1|e^{-\tau/2}$ decreases as long as $\NY<\eta$.  Moreover
Eq.~(\ref{taueq}e) implies that $|\bk|<\epsilon$ remains true as long
$U^2<\epsilon$.  Therefore $\Ype<\epsilon$ as long as $|N|<\eta$.
\endgraf
Next, if $|N|\gg\Ype$ then $\dot N\approx(1-N)N$ due to
Eq.~(\ref{taueq}d) and $r\approx(1-N)^{-1}$ due to Eq.~(\ref{req}),
i.e., $r>1$ and $\dot N>0$ for $N\gg\epsilon$, resp.\ $r<1$ and $\dot
N<0$ for $N\ll-\epsilon$.  Finally, Eq.~(\ref{Wrbound}) implies that $N$
can vanish only when $r<1$.
\end{Prf}

To conclude the argument we analyze what happens to the solutions in the
limit $b\to b_\infty$.
\begin{Prop}{}\label{Plimit}
Given $b_\infty$ and $\eta$ as defined above there exists some
$\delta>0$ such that the solutions with regular origin and
$|b-b_\infty|<\delta$ satisfy:
\begin{enumerate}
\item[1.]
Case~{\bf a} of Lemma~\ref{Lthree} holds if and only if $b=b_\infty$.
There exist continuous functions $\bar\tau(b)<\tau_1(b)$ defined for
$b\ne b_\infty$ such that the same case either~{\bf b} or~{\bf c} holds
for all $b<b_\infty$ and for all $b>b_\infty$ (with
$|b-b_\infty|<\delta$); case~{\bf b} holds in particular for the
globally regular solutions with $n$ zeros of $W$ as $b_n\to b_\infty$
for $n\to\infty$.
\item[2.]
In the limit $b\to b_\infty$ both $\bar\tau$ and $\tau_1-\bar\tau$
diverge.  The part of the solution defined for $\tau<\bar\tau$ converges
for any fixed $\tau$ or $r<1$ to the oscillating solution.  The part
defined for $\tau>\bar\tau$ converges for any fixed $\tau-\tau_1$ or
$r\ne1$ to the exterior, resp.\ interior of the extremal RN solution
with $W\equiv0$ in case~{\bf b}, resp.~{\bf c}.
\end{enumerate}
\end{Prop}
\begin{Prf}{}
\begin{enumerate}
\item[1.]
Since an oscillating solution exists only for finitely many values of
$b$, we can choose $\delta>0$ in Lemma~\ref{Ldelta} such that the
interval $|b-b_\infty|<\delta$ contains only one of them, namely
$b_\infty$.  The existence of $\bar\tau$ and $\tau_1$ for $b\ne
b_\infty$ was shown in Lemma~\ref{Lthree}.  The rest follows from the
continuity of the solutions in $b$.
\item[2.]
The convergence of the solutions follows from the convergence of the
initial data, i.e., quadruples $Y$ at an arbitrary regular point.  The
initial data for any fixed $\tau$ converge to those of the oscillating
solution.  At the same time $\bar\tau$ (with $r(\bar\tau)=1$, resp.\
$N(\bar\tau)=0$) diverges.  On the other hand $Y(\tau_1)$ converges to
$(0,0,\pm\eta,0)$, i.e., to initial data for the exterior or interior of
the extremal RN black hole and $\bar\tau-\tau_1\to-\infty$.  Convergence
for fixed $r$ requires in addition $N\ne0$; given $r\ne1$ this is
satisfied for $b$ sufficiently close to $b_\infty$.
\end{enumerate}
\end{Prf}

Using exactly the same arguments one obtains
\begin{Coro}{}
Analogous results hold true for black hole solutions with any fixed
$r_h<1$ and $W_h$, $W_{hn}$, $W_{h\infty}$ replacing $b$, $b_n$,
$b_\infty$.
\end{Coro}

Having shown the incorrectness of the statements made by Smoller and
Wasserman in \cite{SW} about the limiting solution one may ask for the
source of this error.  Looking at their arguments one finds that they
use Prop.~3.2 of their earlier work \cite{Smoller} in an essential
way.  This proposition is, however, wrong as it stands; its validity
requires the further assumption of a uniformly bounded rotation number
(as made for their Prop.~3.1).  This additional assumption is not
satisfied for the Bartnik-McKinnon family.

\end{document}